\documentclass[aps,prb,superscriptaddress,amsmath,amssymb,floatfix,twocolumn,showpacs]{revtex4}

\pdfoutput=1

\usepackage{amsmath,dcolumn,bm,amsthm,amsfonts,amssymb,tabularx,subfigure}
\usepackage[colorlinks=true,urlcolor=blue,citecolor=blue,linkcolor=blue]{hyperref}
\usepackage{times,bbold,dsfont}
\usepackage{graphicx,graphics,color}

\newcommand{\beq}{\begin{equation}}
\newcommand{\eeq}{\end{equation}}
\newcommand{\bea}{\begin{eqnarray}}
\newcommand{\eea}{\end{eqnarray}}

\begin{document}

\title{Tunable Splitting of the Ground-State Degeneracy in 1D Parafermionic Wires}

\author{Chun~Chen}
\email[]{chen2698@umn.edu}
\affiliation{School of Physics and Astronomy, University of Minnesota, Minneapolis, Minnesota 55455, USA}

\author{F.~J.~Burnell}
\email[]{fburnell@umn.edu}
\affiliation{School of Physics and Astronomy, University of Minnesota, Minneapolis, Minnesota 55455, USA}

\date{\today}

\begin{abstract}

Systems with topologically protected ground-state degeneracies are currently of great interest due to their potential applications in quantum computing. In practise this degeneracy is never exact, and the magnitude of the ground-state degeneracy splitting imposes constraints on the timescales over which information is topologically protected. In this Letter we use an instanton approach to evaluate the splitting of topological ground-state degeneracy in quasi-$1$D systems with parafermion zero modes, in the specific case where parafermions are realized by inducing a superconducting gap in pairs of fractional quantum Hall $($FQH$)$ edges. We show that, like $1$D topological superconducting wires, this splitting has an oscillatory dependence on the chemical potential, which arises from an intrinsic Berry phase that produces interference between distinct instanton tunneling events. These Berry phases can be mapped to chiral phases in a $($dual$)$ quantum clock model using a Fradkin-Kadanoff transformation. Comparing our low-energy spectrum to that of phenomenological parafermion models allows us to evaluate the real and imaginary parts of the hopping integral between adjacent parafermionic zero modes as functions of the chemical potential.

\end{abstract}


\maketitle

The possibility of creating bound states with exotic non-Abelian statistics at the ends of quasi-$1$D systems is an exciting prospect that has been developed both theoretically \cite{Kitaev,Fu,Lutchyn,Oreg,Elliott,Fendley,BarkeshliQi,Clarke,Lindner,Cheng3,Vaezi,ClarkeNatPhys,KlinovajaLoss,Mong1,Jermyn,Barkeshli2} and experimentally \cite{Mourik,Das,Deng,Churchill,Hart,Pribiag,Higginbotham,Finck,NPerge} over the past few years. These bound states are known to generate a topological ground-state degeneracy. In arrays of $1$D systems, states in the resulting low-energy Hilbert space can be entangled and manipulated by braiding processes \cite{AliceaBraiding}, which are $($at least in principle$)$ robust to noise. Such systems have drawn significant interest due to their potential to realize topological quantum computation \cite{FreedmanKitaevLarsenWang,KitaevTQC,NayakReview,NayakDSFreedman,Bonderson}.

\emph{Parafermion} bound states \cite{Fendley,BarkeshliQi,Clarke,Lindner,Cheng3,Vaezi,ClarkeNatPhys,KlinovajaLoss,Mong1,Jermyn,Barkeshli2}, which have $k~(k>2)$ topologically degenerate ground states, are particularly promising for such quantum computing applications: In comparison to Majorana bound states $(k=2)$, parafermion bound states allow for a denser $($albeit non-universal$)$ set of computational gates, and are believed to be intrinsically more robust to environmental noise \cite{Clarke,Lindner,Cheng3,AliceaFendley}. Parafermions are significantly more challenging to produce than their Majorana counterparts: Most proposals entail generating them as defects in $2$D Abelian FQH states, in ways that have yet to be carried out experimentally. However, their non-Abelian statistics are also more complex than for Majoranas, which makes realizing them a particularly exciting prospect.

For bound states confined to the endpoints of a system of finite length $L$, the topological ground-state degeneracy is split by an amount $\Delta E\sim e^{-L/\xi}$, where $\xi$ is proportional to the correlation length in the bulk of the system, causing superpositions of ground states to decohere over time. Though theoretically it is possible to make this splitting as small as required by making $L$ large, there are definite advantages when this splitting can be made small even for modest-length systems. For $1$D topological superconducting wires \cite{DasSarma,Fidkowski} or spin-Hall based superconductors \cite{Fu,Elliott}, this can be achieved by small adjustments in the appropriate chemical potential, since in addition to the exponential fall-off in $L$ the splitting has an oscillatory dependence on the Fermi momentum via $\Delta E \sim e^{-L/\xi}\cos(k_F L)$. The observed oscillations of the splitting of the zero-bias conductance peaks \cite{Churchill} can be viewed as evidence of Majorana bound states \cite{DasSarma,Fidkowski,Zyuzin,Rainis1,Maier}.

Despite the publicity it has had in Majorana systems \cite{Cheng1,Cheng2,Baraban,Lahtinen}, little attention has been given to the ground-state splitting in parafermion systems. Specifically, one might wonder whether these exhibit an analogue of the oscillatory $\cos(k_F L)$ term. In this Letter, we calculate the splitting of the topological ground-state degeneracy in parafermion platforms obtained by inducing superconductivity or ferromagnetism at certain types of edges in FQH states \cite{Clarke,Lindner,Cheng3,Vaezi,Mong1,ClarkeNatPhys}. Following Ref.~\cite{Fidkowski} in the Majorana case, we perform this calculation using a bosonised description of the strongly interacting $1$D system, in which the splitting of the ground-state degeneracy is obtained by an instanton calculation in the resulting sine-Gordon model. Interestingly, as in the Majorana case we do find oscillations in the splitting as functions of chemical potential or applied magnetic field. These intriguing oscillations result from a Berry phase term in the generic sine-Gordon action. We also use our calculation to deduce the magnitude and phase of the hopping coefficients that arise most naturally in parafermion chains, such as those studied by Refs.~\cite{Fendley,Mong1,Mong2,Li,Stoudenmire,Zhuang}.

{\em Model of parafermion zero modes.}---Several groups \cite{Clarke,Lindner,Cheng3,Vaezi,Mong1,ClarkeNatPhys} have suggested that parafermion zero modes can be generated in systems with counterpropagating chiral edges separating two FQH regions with opposite $g$-factors. The edge of interest consists of one right-moving and one left-moving mode with opposite spin polarizations. Two types of electron tunneling processes can open a gap at this edge: Inducing superconductivity $($SC$)$ generates a Cooper-pairing $\Delta(\psi^\dag_{L,\downarrow}\psi^\dag_{R,\uparrow}+\textrm{H.c.})$, while spin backscattering $\mathcal{B}(\psi^\dag_{L,\downarrow}\psi_{R,\uparrow}+\textrm{H.c.})$ can be induced by tunnel-coupling the edge to a ferromagnet $($FM$)$. Any interface between these different types of induced gaps will host parafermion bound states $($Fig.~\ref{fig:fig1}$)$.

\begin{figure}[ht]
\centering
\includegraphics[width=0.413\textwidth]{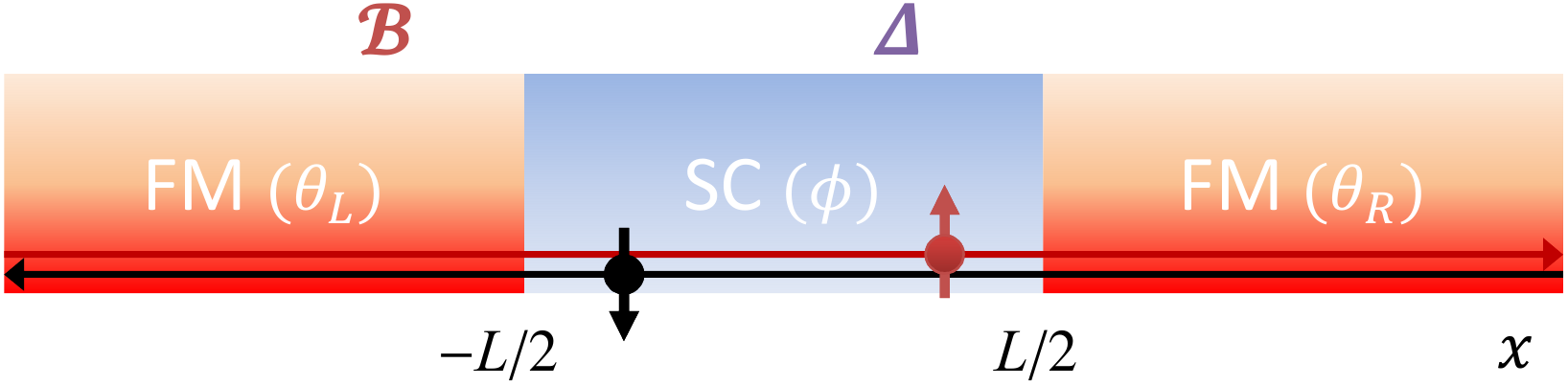}
\caption{\label{fig:fig1} (color online). Schematic spatial profiles of the proximity-induced gaps $\Delta(x)$ and $\mathcal{B}(x)$ for the FM-SC-FM setup.}
\end{figure}

To describe this system, it is convenient to bosonise the two edge modes, representing the right- $($left-$)$ moving electrons as $\psi^\dag_{R/L}\sim\frac{1}{\sqrt{2\pi n\varsigma}}e^{-in\varphi_{R/L}}$, where $\varphi_{R/L}$ are chiral bosonic fields, and $1/n$ is the filling fraction of the corresponding FQH bulk regions. Here $\varsigma$ is related to the inverse energy cutoff $(E_{\textrm{cutoff}})$ of the bosonised theory via $\varsigma\sim v\hbar/E_{\textrm{cutoff}}$, where $v$ is the velocity of the edge modes.

The two backscattering terms are most simply expressed in the basis $\phi=\frac{1}{2}(\varphi_{R,\uparrow}+\varphi_{L,\downarrow})$ and $\theta=\frac{1}{2}(\varphi_{R,\uparrow}-\varphi_{L,\downarrow})$. These non-chiral fields are related to the charge density $\rho_C$ and spin density $\rho_S$ via $\rho_C=\frac{1}{\pi}\partial_x\theta$ and $\rho_S=\frac{1}{\pi}\partial_x\phi$. In this bosonised basis, the two backscattering terms take the form $\Delta(\psi^\dag_{L,\downarrow}\psi^\dag_{R,\uparrow}+\textrm{H.c.})\sim\Delta\sin\!\left(2n\phi\right)$ and $\mathcal{B}(\psi^\dag_{L,\downarrow}\psi_{R,\uparrow}+\textrm{H.c.})\sim\mathcal{B}\sin\!\left(2n\theta\right)$, and the $1$D parafermion system is described by the following Euclidean sine-Gordon action \cite{Clarke,Lindner,Cheng3,Vaezi,Mong1,WenQFT,WenFQHEBosonisation,Giamarchi}:
\bea
S_E\!\!&=&\!\!\!\int\! d\tau dx\{i\hbar\frac{n}{\pi}\partial_x\theta(x,\tau)\partial_{\tau}\phi(x,\tau)\!-\!\frac{\mu(x)}{\pi}\partial_x\theta(x,\tau) \nonumber \\
\!\!&+&\!\!\!\frac{\hbar nv}{2\pi}\!\left(\partial_x\theta(x,\tau)\right)^2\!+\!\frac{\mathcal{B}(x)}{\pi n\varsigma}[\sin{\!(2n\theta(x,\tau))}\!+\!1] \nonumber \\
\!\!&+&\!\!\!\frac{\hbar nv}{2\pi}\!\left(\partial_x\phi(x,\tau)\right)^2\!+\!\frac{\Delta(x)}{\pi n\varsigma}[\sin{\!(2n\phi(x,\tau))}\!+\!1]\}.
\label{SSinG}
\eea
Here $\mu$ represents the chemical potential and $\mathcal{B},\Delta$ are energy gaps induced by spin- and charge-backscattering processes, respectively. The commutation relation $[\phi(x),\theta(x')]=i\frac{\pi}{n}\Theta(x-x')$ dictates that only one of the two possible gapping terms can have a non-vanishing expectation value at a given spatial position. However, if a region where $\Delta\sim\hbar v/\varsigma$ can be sandwiched between two regions where $\mathcal{B}\sim\hbar v/\varsigma$, parafermion bound states arise at the interfaces between them. In the bosonised picture, the resulting topological ground-state degeneracy is manifest in the $2n$ values of $\phi$ for which the sine term is minimized \cite{BarkeshliQi,Clarke,Lindner,Cheng3,Vaezi,Mong1}.

In the following, we consider the FM-SC-FM heterostructure shown in Fig.~\ref{fig:fig1}, on which we take $\mathcal{B}=0~(\Delta,\mu=0)$ in the SC $($FM$)$ region $|x|<L/2~(|x|>L/2)$. In the FM regions $|x|>L/2$, the field $\theta$ is therefore pinned to one of the potential minima, and $\phi$ is strongly fluctuating as required by the commutation relations. As we show in \ref{SupplementaryAct}, under these conditions the FM regions do not contribute to the ground-state energy splitting, and after integrating out $\theta$ we obtain the following effective action for the SC region:
\bea
S_{\phi}\!\!&=&\!\!\!\int^{\frac{T_{\tau}}{2}}_{-\frac{T_{\tau}}{2}} \!\!d\tau\!\! \int^{\frac{L}{2}}_{-\frac{L}{2}}\!\! dx\{\frac{\hbar n}{2\pi v}(\partial_{\tau}\phi(x,\tau))^2+\frac{\hbar nv}{2\pi}(\partial_{x}\phi(x,\tau))^2 \nonumber \\
\!\!&&\!\!+\frac{\Delta}{\pi n\varsigma}[\sin{\!(2n\phi(x,\tau))}+1]+i\frac{\mu(x)}{\pi v}\partial_{\tau}\phi(x,\tau)\}.
\label{effective action}
\eea
In the ground states of this reduced system the $\phi$ field is also approximately pinned at one of the $2n$ inequivalent local minima $\phi_{\textrm{min}}$ of the sine potential; we will take $\Delta$ to be sufficiently large that these low-energy states are well separated from the rest of the spectrum. The fluctuation-induced splitting between the $2n$ otherwise degenerate ground states is then determined by the amplitude for tunneling between adjacent local minima.

The last term in Eq.~(\ref{effective action}) plays the role of a topological Berry phase term $S_{\textrm{B-p}}$, contributing to the net action only for field configurations which start and end at different values of $\theta$ $($i.e. only for instantons$)$. It introduces oscillations in the splitting of the ground-state degeneracy as the chemical potential $\mu$ is varied. To the best of our knowledge, $S_{\textrm{B-p}}$ was not included in previous studies of instantons in the bosonised periodic sine-Gordon model, which considered the case $\mu=0$ \cite{Fidkowski,Bajnok,Giamarchi}.

{\em Instanton calculation of level splitting.}---For the sine-Gordon model described by Eq.~(\ref{effective action}), the minima of the potential term are at $\phi_{\textrm{min}}=-\frac{\pi}{4n}+\frac{j\pi}{n}$ where $j=0,1,\ldots,2n-1$, with the $j$-th ground state denoted by $|j\rangle$. The classical soliton solution interpolating between vacua at $j\frac{\pi}{n}\!-\!\frac{\pi}{4n}$ and $(j\pm 1)\frac{\pi}{n}\!-\!\frac{\pi}{4n}$ has the form \cite{FK,Skyrme1,Skyrme2}:
\beq
\phi_{sol}(\tau)=-\frac{\pi}{4n}+j\frac{\pi}{n} \pm \frac{2}{n}\arctan{\![e^{\omega(\tau-\tau_0)}]},
\label{eq:phi_sol}
\eeq
where $\omega\!=\!2\sqrt{\Delta v/(\hbar\varsigma)}\!>\!0$. To a good approximation, we may neglect spatial variations in the instanton solution due to boundary effects (see Appendix \ref{SupplementaryAct}).

Following Refs.~\cite{Bajnok,Munster,Polyakov,Coleman,Vainshtein}, the amplitudes for the one-instanton and one-anti-instanton processes are:
\begin{align}
\langle j|e&^{-H_{\phi}T_{\tau}/\hbar}|j\mp1\rangle_{o.i.} \nonumber \\
&=\left(\mathcal{N}e^{-\tilde{\omega}T_{\tau}/2}\right)\!\!\left(\sqrt{\frac{v}{L}}\sqrt{\frac{S_0}{2\pi\hbar}}e^{-S_0/\hbar\mp i\gamma}\right)\!\sqrt{\omega}T_{\tau}.
\label{eq_o_i}
\end{align}
Here $S_0\!+\!(-)i\hbar\gamma$ is the effective action of an instanton $($anti-instanton$)$ tunneling event $($with small quantum corrections omitted, as discussed in Appendix \ref{SupplementaryInst}$)$, where the imaginary contribution $\gamma$ stems from $S_{\textrm{B-p}}$, and $\hbar\tilde{\omega}/2$ stands for the zero-point energy of the $1$D harmonic oscillator. In terms of the parameters in Eq.~(\ref{effective action}), we have $S_0\!=\!\frac{2\hbar\omega L}{n\pi v},~\gamma\!=\!\frac{\mu L}{\hbar nv}$. As anticipated, the amplitude for tunneling vanishes exponentially with the wire length $L$ provided the bulk is gapped $($i.e. $\Delta\neq0)$. More significantly, we now perceive the importance of the Berry phase term, which contributes a different net phase to the amplitude for instanton and anti-instanton processes.

\begin{figure}[ht]
\centering
\includegraphics[width=0.424\textwidth]{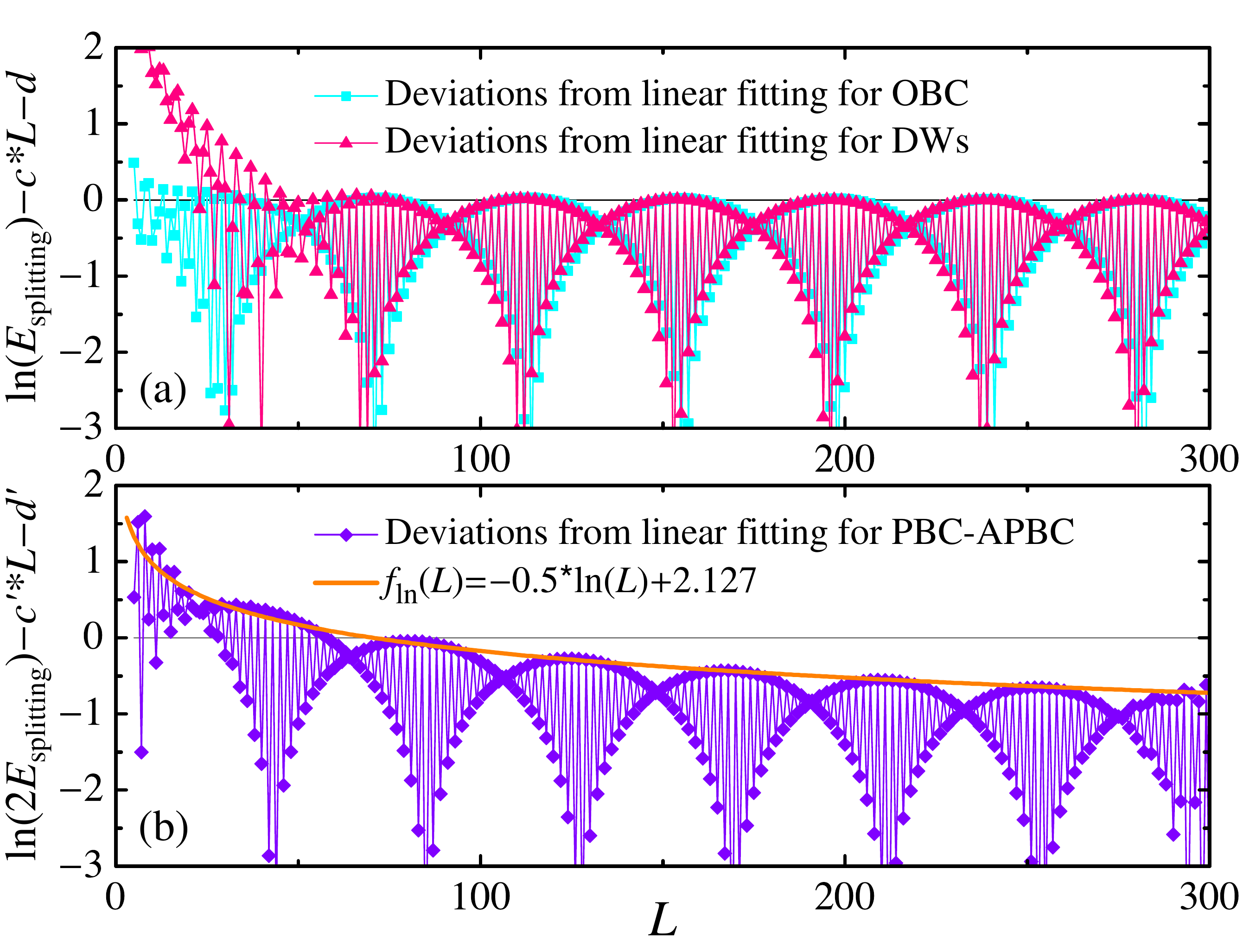}
\caption{\label{fig:fig2} (color online). Numerical results for the \emph{oscillatory} zero-energy splitting as a function of $L$ in the $1$D nanowire models hosting Majorana bound states \cite{Kitaev,Lutchyn,Oreg}, for (a) open and domain-wall BCs, and (b) periodic versus antiperiodic BCs $($PBC-APBC$)$. $($Recall that for a Majorana SC ring, the ground state with PBC $($APBC$)$ is a parity-even $($parity-odd$)$ state.$)$ The term linear in $L$ has been subtracted off in both cases, leaving a result independent of $L$ in (a), and depending logarithmically on $L$ in (b).}
\end{figure}

The prefactor $\mathcal{K}=\sqrt{v\omega S_0/(2\pi\hbar L)}$ is the Fredholm determinant describing Gaussian fluctuations about the saddle-point solution (\ref{eq:phi_sol}). This determinant is sensitive to spatial fluctuations in the SC region, and its scaling with $L$ is sensitive to the choice of boundary conditions $($BCs$)$. As shown in Appendix \ref{SupplementaryInst}, for Neumann-type BCs appropriate to the setup shown in Fig.~\ref{fig:fig1}, this prefactor is independent of $L$, in agreement with existing work on Majorana bound states \cite{Zyuzin,DasSarma,Shivamoggi}. For the energy splitting in periodic BCs relative to antiperiodic ones, on the other hand, the prefactor contains an extra factor of $\frac{1}{\sqrt{L}}$ \cite{Bajnok}. Fig.~\ref{fig:fig2} compares this prediction to numerical values for the energy splitting in the Majorana nanowire for both BCs.

Armed with one-instanton solutions, the total instanton contribution is obtained using a dilute instanton gas approximation \cite{Coleman}, which gives:
\begin{align}
\langle j_+|e^{-H_{\phi}T_{\tau}/\hbar}&|j_-\rangle=\mathcal{N}\int^{2\pi}_0\frac{d\zeta}{2\pi}e^{i\zeta(j_--j_+)} \nonumber \\
\times\exp&{\left[-\frac{\tilde{\omega}}{2}T_{\tau}+2\mathcal{K}T_{\tau}e^{-S_0/\hbar}\cos{\!\left(\zeta-\gamma\right)}\right]}.
\label{TotalEnergy}
\end{align}
The cosine term can be viewed as arising from interference between instanton and anti-instanton trajectories, for which $S_{\textrm{B-p}}$ has the opposite phase.

In the presence of instanton tunneling events, the eigenstates of $H_{\phi}$ are therefore Bloch-wave-like states of the form: $|\zeta\rangle\!\propto\!\left(\frac{1}{2\pi}\right)^{\frac{1}{2}}\sum_je^{-i\zeta j}|j\rangle$. Imposing $\langle j_+\pm 2n|e^{-H_{\phi} T_{\tau}/\hbar}|j_-\pm 2n\rangle\!=\!\langle j_+|e^{-H_{\phi} T_{\tau}/\hbar}|j_-\rangle$ to account for the fact that only $2n$ of these minima are physically distinct forces $\zeta$ to take on the $2n$ discrete values $\frac{\pi}{n}q_{\phi}$, with $q_{\phi}=0,1,\ldots,2n-1$, which gives the $2n$ states:
\beq
|q_{\phi}\rangle\propto\left(\frac{1}{2n}\right)^{\frac{1}{2}}\sum^{2n-1}_{j=0}e^{-i\frac{\pi}{n}q_{\phi} j}|j\rangle.
\label{eqn_q}
\eeq
From Eq.~(\ref{TotalEnergy}), the energies of these states are, up to an overall constant,
\beq
E(q_{\phi})=-\frac{2\hbar\omega}{\pi}\sqrt{\frac{1}{n}}e^{-S_0/\hbar}\cos{\!\left(\frac{\pi}{n}q_{\phi}-\frac{\mu L}{\hbar nv}\right)},
\label{eqn_E}
\eeq
where $q_\phi$ is the generalised charge parity conjugate to the discrete, compact variable $\phi_\textrm{min}$. This form agrees with the general result of Ref.~\cite{BondersonPRL} for the energies of anyon pairs:
\beq
E(q_{\phi})=\sum_{a=0}^{2n-1}\left(\Gamma_a[F^{\alpha a\alpha }_{q_{\phi}}]_{\alpha\alpha}+\textrm{H.c.}\right),
\eeq
where $\alpha$ denotes the parafermion bound states, and $F$ is a part of the topological data. For the case at hand the possible choices of $F$ are described in Ref.~\cite{LongQPaper}; taking $[F^{\alpha a\alpha}_{q_{\phi}}]_{\alpha\alpha}\!=\!e^{i(\pi/n)[(a\cdot q_{\phi})~\textrm{mod}~2n]}$ and $\Gamma_a\!=\!-\hbar\mathcal{K}e^{-S_0/\hbar-i\gamma}\delta_{a,1}$ recovers the form (\ref{eqn_E}) for the energies of these ground states.

It is instructive to check Eq.~(\ref{eqn_E}) for the case of Majorana fermions $(n=1)$, where the splitting can be calculated directly from a quadratic fermion Hamiltonian \cite{DasSarma,Elliott,Cheng1,Cheng2}. The relevant calculation in the quantum Hall systems described here can be carried out as for the nanowire case \cite{Zyuzin,DasSarma} -- see Ref.~\cite{Shivamoggi} and Appendix \ref{SupplementarySplit}. The result is
\beq
E_{q}=-C\frac{\Delta\mathcal{B}}{\Delta+\mathcal{B}}e^{-\frac{\Delta L}{\hbar v}}\cos{\!\left(\pi q-\frac{\mu L}{\hbar v}\right)},~~q=0,1,
\eeq
where $C$ is a constant of order unity. The coefficient of the decaying exponential differs from Eq.~(\ref{eqn_E}), since for $n=1$, $S_0/\hbar\!=\!\left(\frac{\Delta L}{\hbar v}\right)\frac{4}{\pi}\sqrt{\hbar v/(\Delta\varsigma)}$. However, in the instanton calculation of the exponential term we neglect all modes above the gap set by $\Delta$; hence the cutoff energy $\hbar v/\varsigma$ should not be much larger than $\Delta$. Further, the tunneling process requires the presence of virtual fluctuations up to an energy of approximately $\Delta$, so the cutoff energy should also not be much smaller than $\Delta$. The factor $\frac{4}{\pi}\sqrt{\hbar v/(\Delta\varsigma)}$, which parameterizes the difference in the exponential decay lengths from the two calculations, is therefore a constant of order $1$. Parallel reasoning applies to the comparison of the prefactors. Remarkably, the argument of the cosine term agrees exactly with our instanton calculation, suggesting that this oscillatory dependence on $\mu$ is insensitive to the BCs and to the various approximations being made.

We note that the preceding analysis also applies to the case $\mathcal{B}>0,\Delta=0$ by taking $\phi\rightleftharpoons\theta$ and replacing the chemical potential term with a magnetic field term of the form $-\frac{h}{\pi}\partial_x\phi$. Therefore the oscillatory dependence of the ground-state energy splitting on chemical potential or magnetic field is a relatively ubiquitous feature of parafermion zero modes.

{\em Hopping in parafermion chains.}---One interesting application of our calculation is that it allows us to infer the phase of intrawire parafermion hopping terms. This is of particular interest as chains of coupled parafermions can be used to generate even more exotic topological phases \cite{Mong1,Stoudenmire}.

In the setup we consider, the parafermion bound states can be described by operators $\alpha^{(\dag)}_L,\alpha^{(\dag)}_R$ which annihilate $($create$)$ parafermion zero modes at the left and right endpoints of the SC region, respectively, and satisfy the relations
\beq
\alpha^{2n}_{L/R}=1,~\alpha^{\dagger}_{L/R}=\alpha^{2n-1}_{L/R},~\mbox{and}~\alpha_L\alpha_R=\alpha_R\alpha_L e^{i\frac{\pi}{n}},
\eeq
which are sufficient to ensure that these bound states have \lq\lq parafermionic'' non-Abelian statistics \cite{BarkeshliQi,Clarke,Lindner,Cheng3,Vaezi}.

In terms of the bosonised fields $\theta$ and $\phi$, we have $\alpha^{\dagger}_{L}\alpha_{R}=e^{i\frac{\pi}{n}(q-1/2)}$, where $q=\frac{n}{\pi}(\theta(\frac{L}{2})-\theta(-\frac{L}{2}))$ is the total charge in the SC segment modulo $2$ \cite{Clarke,Lindner,Cheng3,Vaezi}. From the commutation relation between $\phi$ and $\theta$, it follows that
\beq
(\alpha^{\dagger}_R\alpha_L)\phi(\alpha^{\dagger}_L\alpha_R)=\phi+\frac{\pi}{n},~(\alpha^{\dagger}_L\alpha_R)\phi(\alpha^{\dagger}_R\alpha_L)=\phi-\frac{\pi}{n}, \nonumber
\eeq
which carrying out precisely the tunneling processes whose matrix elements we have just evaluated.

The low-energy Hamiltonian describing the parafermion tunneling between the two endpoints is therefore
\beq
H_A=t\alpha^{\dagger}_{L}\alpha_{R}+t^*\alpha^{\dagger}_{R}\alpha_{L}. \nonumber
\label{eq:H_A}
\eeq
Its eigenstates are labeled by an integer $q=0,1,2,\ldots,2n-1$, and satisfy \cite{Clarke,Lindner,Cheng3,Vaezi}
\beq
\alpha^{\dagger}_{L}\alpha_{R}|q\rangle=-e^{i\frac{\pi}{n}(q-\frac{1}{2})}|q\rangle,~\alpha^{\dagger}_{R}\alpha_{L}|q\rangle=-e^{-i\frac{\pi}{n}(q-\frac{1}{2})}|q\rangle.
\label{eq:para_splitting}
\eeq
Note that Eqs.~(\ref{eqn_q}) and (\ref{eq:para_splitting}) together also fix the phase associated with the action of the parafermion hopping term on the eigenstates of $\phi$: $\alpha^{\dagger}_L\alpha_R|j\rangle\!=\!-e^{-i\frac{\pi}{2n}}|j+1\rangle,~\alpha^{\dagger}_R\alpha_L|j\rangle\!=\!-e^{+i\frac{\pi}{2n}}|j-1\rangle$. The corresponding energies---which are precisely the energies that we have just obtained with our instanton calculation---are:
\beq
E(q)=-2\sqrt{t^2_\Re+t^2_{\Im}}\cos{\!\left[\frac{\pi}{n}\left(q-\frac{1}{2}\right)+\vartheta\right]},
\label{eq:splitting_parafermion}
\eeq
where we have defined $\tan{\vartheta}=t_{\Im}/t_{\Re}$.

Comparing Eqs.~(\ref{eqn_E}) and (\ref{eq:splitting_parafermion}) allows us to constrain $t_\Re$ and $t_\Im$. For Majorana fermions $($i.e. $n=1)$, there is an additional constraint: Since $\alpha^\dag_{R/L}=\alpha_{R/L}$, the two terms in Eq.~(\ref{eq:H_A}) are not independent. This forces $t_{\Re}=0$ $($i.e. $\vartheta=\pm\pi/2)$, and $t_{\Im}=\pm\hbar\mathcal{K}e^{-S_0/\hbar}\cos{\![\mu L/(\hbar nv)]}$. For $n>1$ there is no such a constraint; however, in these cases matching the eigenvalues of both $H$ and the operator $\alpha^{\dagger}_L\alpha_R=e^{i\left(\theta(\frac{L}{2})-\theta(-\frac{L}{2})-\pi/(2n)\right)}$ fixes $\vartheta$, such that
\begin{align}
t_{\Re}&=\pm\hbar\mathcal{K}e^{-S_0/\hbar}\cos{\!\left[\pi/(2n)-\mu L/(\hbar nv)\right]}, \label{eqn:t_R} \\ t_{\Im}&=\pm\hbar\mathcal{K}e^{-S_0/\hbar}\sin{\!\left[\pi/(2n)-\mu L/(\hbar nv)\right]}. \label{eqn:t_I}
\end{align}
Using the analogous approach for an SC-FM-SC system $($with $\mu\Rightarrow h, \Delta\Rightarrow\mathcal{B})$ gives the analogous conclusion.

It is worth stressing that even at vanishing $\mu$ $($or~$h)$, for $n>1$ the hopping parameter $t$ is complex with $\vartheta=\pi/(2n)$. This suggests that the proposal for universal quantum computing by manufacturing Fibonacci anyons in coupled parafermion chains \cite{Mong1,Mong2,Stoudenmire,Zhuang} is better achieved in systems without finite chemical potential or magnetic field. More specifically, the $\mu$-dependent contribution to $\vartheta$ in Eqs.~(\ref{eqn:t_R}) and (\ref{eqn:t_I}) corresponds to a \emph{chiral} phase \cite{Fendley2,Baxter,AuYangPerk,FootnoteAdd3} in the quantum clock model. To be concrete, a system of $2N$ tunnel-coupled parafermion zero modes is dual $($via the Fradkin-Kadanoff mapping \cite{FradkinKadanoff,Fendley,Zhuang}$)$ to an $N$-site chiral quantum clock chain, where with appropriate conventions hopping across a SC $($FM$)$ region maps to the transverse field $($ferromagnetic clock$)$ coupling. Under duality, the phases of the parafermion hopping terms $t$ map to a chiral phase of $\pm e^{i\frac{hL}{\hbar nv}}$ for the ferromagnetic clock coupling, as well as a chiral phase of $\pm e^{i\frac{\mu L}{\hbar nv}}$ for the transverse field term. Notice that the oscillatory dependence of the ground-state degeneracy splitting on the chiral phases and the system's size has been observed numerically for these chiral clock systems \cite{Zhuang} recently.

In summary, our non-perturbative calculation shows that it is possible, in principle, to tune the magnitude of the ground-state splitting in parafermion systems $($as well as the phase of the parafermion hopping parameter, for $n>1)$ by means of a chemical potential or an external magnetic field due to interference between distinct instanton trajectories resulting from a topological term in the effective action. Because the period of the resulting oscillations is given by $\mu L/(\hbar n v)$ $($or $hL/(\hbar n v))$, this splitting can be fine-tuned with relatively small changes in $\mu$ $($or $h)$. As for Majoranas, we anticipate that this fact will be both of practical use to achieve quantum-coherent systems, and a potential signature of the existence of parafermions in these systems. Finally, our results might also be applicable to the spin-unpolarized $\nu=2/3$ FQH heterostructures proposed by Refs.~\cite{Mong1,ClarkeNatPhys}.

{\em Acknowledgements}: We thank J. Alicea, C. L. Kane, S. Ryu, M. D. Schulz, Q. Shi, and M. Zudov for useful discussions. We also thank the anonymous Referees whose suggestions and questions improve the manuscript a lot. FJB is supported by NSF-DMR 1352271 and by the Sloan foundation FG-2015-65927.

\begin{widetext}

\appendix

\section{The reduced effective action for instanton configurations} \label{SupplementaryAct}

Here we justify the effective action $S_{\phi}$ which is our starting point for studying instanton effects in the main text. Specifically, we will show that, for a short SC region sandwiched between two long FM regions, the effect of instantons is limited to the SC region. We will use this to derive an effective action in the SC region alone, with appropriate boundary conditions at the boundaries of this region.

To address the details of this problem, we must be slightly more precise in what we mean by FM and SC regions. We consider the FM-SC-FM heterostructure where the edge is gapped by FMs when $x\leq-\frac{L+\delta}{2}$ and $x\geq\frac{L+\delta}{2}$, while in the interval $-\frac{L-\delta}{2}\leq x\leq\frac{L-\delta}{2}$, the edge is gapped by SC. To simplify the calculation, we approximate these gaps as step functions $($see Fig.~\ref{fig:fig1}$)$. The commutation relations between $\theta$ and $\phi$ make it impossible to induce both kinds of gaps at the same spatial position; therefore there is necessarily a small domain wall $($DW$)$ region between FM and SC regions, whose width we take to be $\delta\ll L$.

Our derivation proceeds in three steps. First, we will argue that, provided the FM gap is large, the instanton configurations of interest have essentially no effect on physics in the FM region. This allows us to restrict our attention to the effective action in the SC and DW regions, with appropriate boundary conditions at each domain wall. Second, we integrate out $\theta$ in the SC and DW regions to obtain an effective action for the field $\phi$. Finally, we argue that the specified boundary conditions mean that the contribution of the DW region to the tunneling amplitude is also essentially independent of the instanton configuration, allowing us to study the ground-state splitting using an effective action for the SC region alone.

\begin{figure*}[ht]
\centering
\includegraphics[width=0.6\textwidth]{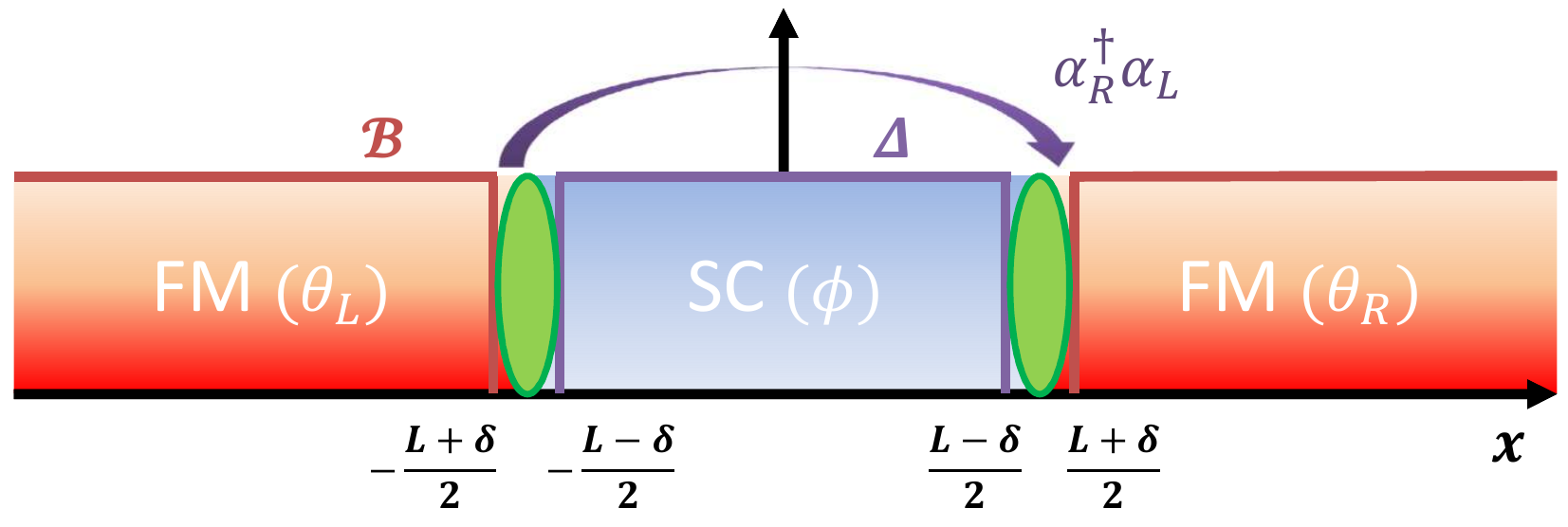}
\caption{\label{fig:fig1} (color online). The spatial profiles of the proximity-induced gaps $\Delta(x)$ and $\mathcal{B}(x)$ for the FM-SC-FM setup. The two gapped regions are necessarily separated by a domain wall region of width $\delta$ that accommodates a parafermionic zero mode denoted by the green ellipse.}
\end{figure*}

\textbf{The FM region.} First, we will argue that the impact of the FM region on the ground-state splitting can be captured simply by choosing appropriate boundary conditions for the fields $\theta,\phi$ at $x=\pm\frac{L+\delta}{2}$. Essentially, this is because the instanton profile vanishes for $|x|>\frac{L+\delta}{2}$, and hence the path integral on these regions is indifferent to instanton effects.

To see why this must be so, we begin by considering the limit $\frac{2n\mathcal{B}}{\pi\varsigma}\rightarrow\infty$. In this limit $\theta$ is exactly frozen in each FM region to one of its $2n$ minimal values, and its spatial and temporal derivatives in this region vanish. Since $[\frac{n}{\pi}\partial_x\theta(x),\phi(x)]=i$, pinning $\partial_x\theta\equiv0$ in this region forces $\phi$ to be evenly distributed on the interval $[0,2\pi)$, irrespective of the behaviour of $\phi$ for $|x|<\frac{L+\delta}{2}$. Hence in the limit that $\theta$ and its derivatives are perfectly pinned in the FM regions, their contribution to the tunneling amplitude is necessarily independent of the instanton configuration in the SC and DW regions. Hence in this limit, we can replace the FM region with free boundary condition for $\phi(\pm(L+\delta)/2)$, and a sum over all possible pinned values of $\theta$ at each of these boundaries. $($This sum ensures that the system can access all possible values of $q$.$)$

In other words, for $\frac{2n\mathcal{B}}{\pi\varsigma}\rightarrow\infty$ the tunneling amplitude for the entire system can be expressed:
\beq
\mathcal{I}=\sum_{\theta_{R},\theta_{L}}\int d\phi_1 d\phi_2 Z_{\textrm{FM}}[\theta_{L},\phi_1]\mathcal{T}[\theta_{L},\theta_{R},\phi_1,\phi_2]Z_{\textrm{FM}}[\theta_{R},\phi_2],
\label{PartFactored}
\eeq
where the summation is over all values of $\theta_{L},\theta_{R}\in\frac{\pi}{n}\{0,1,\ldots,2n-1\}$. Here we have defined $\phi\left(-(L+\delta)/2\right)=\phi_1,\phi\left((L+\delta)/2\right)=\phi_2$. Note that we must integrate over all values of $\phi_1,\phi_2$, since $\partial_x\theta$ is pinned at this boundary.

From the form of the action in the FM regions $($which depends only on derivatives of $\phi$, with no potential term$)$, it is evident that $Z_{\textrm{FM}}$ will be independent of the particular values of $\phi_{1,2}$, and also of the choice of $\theta_{L,R}$ within the set of minima of the potential. Therefore any impact from instanton configurations must be contained in the tunneling amplitude $\mathcal{T}[\cdots]$, which describes the SC and DW regions.

Hence provided $\mathcal{B}$ is sufficiently large that fluctuations in $\theta$ in the FM region can be neglected, the FM region cannot contribute to the instanton transition amplitude. Including small fluctuations, to account for the fact that $\partial_x\theta$ is not exactly pinned in the FM region, we would find that the boundary values of $\phi$ need not be perfectly uniformly distributed, making some \lq\lq leaking'' of the instanton configuration into the FM region possible. However, provided the fluctuations in $\theta$ are small $($i.e. the FM gap is large$)$, the distribution of $\phi$ must be wide. The finite correlation length in this region also ensures that any correlation with the instanton solution occurs over a finite $($and small$)$ spatial extent. Therefore though the impact of the instanton configuration on the FM region in this case is not strictly $0$, we will neglect it relative to the remaining terms, an approximation that is valid provided the FM gap remains large. Including such terms could result in corrections of order $1$ of the magnitude of the instanton transition amplitude, but will not contribute to the Berry phase term in the action provided $\mu=0$ in the FM regions.

\textbf{Effective action in the SC and DW regions.} Based on the previous discussion, to calculate the ground-state splitting we may focus on the region $|x|<\frac{L+\delta}{2}$, where the potential term for $\theta$ vanishes. The boundary conditions on this region are that $\phi$ is free at both spatial boundaries, and that $\theta\left(-\frac{L+\delta}{2}\right)=\theta_{L},~\theta\left(\frac{L+\delta}{2}\right)=\theta_{R}$, where the transition amplitude contains a sum over the possible values $\frac{n}{\pi}\theta_{R,L}=0,1,\ldots,2n-1$. Recall that the total charge in the SC segment is fixed by $q_{\phi}/n=\frac{1}{\pi}\left(\theta\left(\frac{L+\delta}{2}\right)-\theta\left(-\frac{L+\delta}{2}\right)\right)=0,1/n,\ldots,(2n-1)/n$ $($mod $2)$; hence this sum ensures that all possible charges are included in the path integral.

Within this region, the potential term for $\theta$ vanishes, and we may complete the square in the path integral to obtain:
\begin{align}
S=\int d\tau\int_{|x|<\frac{L+\delta}{2}}dx&\left\{\frac{\hbar nv}{2\pi}\left[\partial_x\theta(x,\tau)-\frac{\mu(x)}{\hbar nv}+\frac{i}{v}\partial_{\tau}\phi(x,\tau)\right]^2\right. \nonumber \\
&+\frac{\hbar n}{2\pi v}\left(\partial_{\tau}\phi(x,\tau)\right)^2+\frac{\hbar nv}{2\pi}\left(\partial_{x}\phi(x,\tau)\right)^2+\frac{\Delta(x) }{\pi n\varsigma}\left[\sin{\!(2n\phi(x,\tau))}+1\right] \nonumber \\
&+\left.i\frac{\mu(x)}{\pi v}\partial_{\tau}\phi(x,\tau)-\frac{(\mu(x))^2}{2\pi\hbar nv}\right\}.
\label{S_sc}
\end{align}
The second line in Eq.~(\ref{S_sc}) is the standard sine-Gordon action $S_{\textrm{s-G}}$, and the first term in the third line is the newly-derived Berry phase term $S_{\textrm{B-p}}$. Shifting $\theta$ and performing the Gaussian integral leads to the effective action given in the main text:
\begin{align}
S_{\phi}&=\int d\tau\int_{|x|<\frac{L+\delta}{2}}dx\left\{\frac{\hbar n}{2\pi v}\left(\partial_{\tau}\phi(x,\tau)\right)^2+\frac{\hbar nv}{2\pi}\left(\partial_{x}\phi(x,\tau)\right)^2+\frac{\Delta(x)}{\pi n\varsigma}\left[\sin{\!(2n\phi(x,\tau))}+1\right]+i\frac{\mu(x)}{\pi v}\partial_{\tau}\phi(x,\tau)\right\},
\label{S_sc2}
\end{align}
where we have dropped the constant term.

In the main text, we ignore the domain wall region, and treat the classical instanton field as position-independent. Relaxing this approximation will result in corrections of order $\delta/L$ to the classical instanton action. In particular, the contribution of time derivatives of $\phi$ in this region is clearly small in $\delta/L$. Further, after averaging over all boundary values at $|x|=\frac{L+\delta}{2}$, the set of differences of $\phi$ across the spatial extent of the DW region is independent of the pinned value of $\phi$ in the SC region $($and its time derivative$)$ -- i.e. the distribution of $\partial_x\phi$ in the DW region is insensitive to the instanton configuration, and hence does not contribute to the instanton transition amplitude.

\textbf{Boundary terms from integrating by parts.} To obtain the action $S_E$ given in the main text, we have integrated a term of the form $\partial_\tau\theta\partial_x\phi$ by parts. Here we verify that the resulting boundary terms do not contribute to the effective action of the instanton. To show this, we use periodic boundary conditions in time for the fluctuating components of our fields.

The boundary terms in question are: $\int dx\left[\left(i\theta(x,\tau)\partial_x\phi(x,\tau)\right)\big\vert^{T_{\tau}/2}_{-T_{\tau}/2}\right]-\int d\tau\left[\left(i\theta(x,\tau)\partial_{\tau}\phi(x,\tau)\right)\big\vert^{R}_{L}\right]$. We begin with the first term. In the SC region, we may separate $\phi=\phi_{sol}+\eta$, where $\eta(x,T_{\tau}/2)=\eta(x,-T_{\tau}/2)$ and $\partial_x\phi_{sol}=0$. In the FM region $\phi\equiv\eta$, and since there are no instantons in $\theta$ the configurations at $\pm T_{\tau}/2$ are identical. In the DW region the field does vary spatially, but this variation is not due to $\phi_{sol}$, but rather due to the fluctuations $($in other words, the distribution of $\partial_x\phi$ in the DW is time-independent, as emphasized above$)$. Therefore the first term vanishes.

For the second term, $($which we compute for the whole system, not only the region relevant to the instanton action$)$, we obtain:
\begin{align}
\int d\tau\left[\left(i\theta(x,\tau)\partial_{\tau}\phi(x,\tau)\right)\big\vert^{x=\infty}_{x=-\infty}\right]=\theta_R\int d\tau\left[i\partial_{\tau}\phi(\infty,\tau)\right]-\theta_L\int d\tau\left[i\partial_{\tau}\phi(-\infty,\tau)\right] \nonumber \\
=\theta_R(\phi(\infty,T_{\tau}/2)-\phi(\infty,-T_{\tau}/2))-\theta_L(\phi(-\infty,T_{\tau}/2)-\phi(-\infty,-T_{\tau}/2))=0, \nonumber
\end{align}
where we have used the fact that since each FM region is semi-infinite, there are no instantons in $\theta$. The last equality follows from the fact that in the FM region $\phi$ consists only of fluctuations, which obey periodic boundary conditions in $($imaginary$)$ time.

Thus both of these boundary terms do indeed vanish.

{\bf Shifting $\theta$.} Here we present the details of the shift in $\theta$, to confirm that the resulting integral is independent of the configuration of $\phi$. For a one-instanton contribution, we have $\phi(x,\tau)=\phi_{sol}(x,\tau)+\eta(x,\tau)$, where the fluctuation field $\eta$ obeys $\eta(x,\tau\!=\!\pm T_{\tau}/2)\!=\!0$, and $\phi_{sol}$ is the classical instanton solution, which we argued above vanishes at $x=\pm\frac{L+\delta}{2}$. We must now evaluate
\beq
\int\mathcal{D}\theta (x,\tau)\exp{\left\{-\frac{nv}{2\pi}\int d\tau\int_{|x|<\frac{L+\delta}{2}}dx[\partial_x\theta(x,\tau)-\frac{\mu(x)}{\hbar nv}+\frac{i}{v}\partial_{\tau}\eta(x,\tau)+\frac{i}{v}\partial_{\tau}\phi_{sol}(x,\tau)]^2\right\}}.
\eeq

Since the region in question is finite, one might worry that such a shift will result in different boundary conditions for $\theta$ and $\tilde{\theta}$, either in space or in time, leading to a different set of eigenvalues even for the same differential operator. However, the boundary conditions are not affected by $\phi_{sol}$ or $\mu$. First, both of these quantities vanish at the spatial boundary. Second, in the limit of interest $T_{\tau}\rightarrow\infty$, $\partial_\tau\phi_{sol}$ vanishes at the time-like boundaries, and $\mu$ is time independent. This leaves the fluctuation field $\eta$, which obeys free rather than fixed BCs at the spatial boundaries $($but the same BCs in time$)$. This change in spatial BCs between $\theta$ and $\tilde{\theta}$ does not, however, result in any additional dependence on $\eta$ in the functional integral.

To perform the desired shift, we Fourier transform all fields by using $\theta(x,\tau)=\frac{1}{T_{\tau}L}\sum_{k,i\omega}\theta(k,i\omega)e^{ikx}e^{-i\omega\tau}$ $($and similarly for other fields$)$, where $\omega$ are real, discrete frequencies, then shift the Fourier components $\theta(k,i\omega)$ to absorb the chemical potential and the soliton terms into the Gaussian integral of $\theta$. This gives: $\tilde{\theta}(k,i\omega)=\theta(k,i\omega)-\frac{i\mu(k)T_{\tau}\delta_{\omega,0}}{\hbar nvk}+\frac{i\omega}{kv}\left(\eta(k,i\omega)+\phi_{sol}(k,i\omega)\right)$, where $\delta_{\omega,0}$ is the Kronecker $\delta$-function. The resulting Gaussian integral over $\theta$ is independent of the specific form of both $\phi_{sol}$ and $\eta$, leading to the effective action (\ref{S_sc2}).

\section{Splitting of ground-state energy in spin-Hall edge: A wavefunction-overlap calculation} \label{SupplementarySplit}

As an important check, let us discuss the Majorana zero modes at the edge of a $2$D quantum spin-Hall system which is in close proximity to an $s$-wave superconductor and two ferromagnetic insulators. We assume that the right movers carry spin up and the left movers carry spin down. The Bogoliubov--de Gennes $($BdG$)$ Hamiltonian for the edge states is as follows,
\begin{equation}
\mathcal{H}(x)=-iv_F\tau_z\sigma_z\partial_x-\mu(x)\tau_z-\Delta(x)\tau_x+\mathcal{B}(x)\sigma_x.
\label{eq:realhami}
\end{equation}
The Pauli matrices $\vec{\tau}$ and $\vec{\sigma}$ act on the particle-hole and spin spaces, respectively. To focus our discussion, we consider the FM-SC-FM setup shown in Fig.~\ref{fig:fig1}. Neglecting the width of the domain wall region, this gives the following spatial profiles for the chemical potential and the proximity-induced gapping terms: $\mu(x)=\mu\Theta(x+\frac{L}{2})\Theta(-x+\frac{L}{2}),~\Delta(x)=\Delta\Theta(x+\frac{L}{2})\Theta(-x+\frac{L}{2}),~\mathcal{B}(x)=\mathcal{B}(1-\Theta(x+\frac{L}{2})\Theta(-x+\frac{L}{2}))$. Here the potential strengths $\Delta,\mathcal{B}\gg|\mu|\neq0$ and $\Theta(x)$ is the Heaviside step function. We also assume that the chemical potential term can be generated by locally gating the edge.

With the linear dispersion relation in Eq.~(\ref{eq:realhami}), analytical solutions for the Majorana bound states can easily be obtained in the limit that the two bound states are infinitely far apart. Specifically, if we divide $\mathcal{H}(x)$ into the left and right parts for $x\in(-\infty,\infty)$, then $\mathcal{H}(x)=\mathcal{H}_{L}(x)+\mathcal{U}_{R}(x)=\mathcal{H}_{R}(x)+\mathcal{U}_{L}(x)$, where
\beq
\mathcal{H}_{L}(x)=\left(
\begin{array}{cccc}
 -iv_F \partial_x-\mu_L(x) & \mathcal{B}_L(x) & -\Delta_L(x) & 0  \\
 \mathcal{B}_L(x) & iv_F \partial_x-\mu_L(x) & 0 & -\Delta_L(x)  \\
 -\Delta_L(x) & 0 & iv_F \partial_x+\mu_L(x) & \mathcal{B}_L(x)  \\
 0 & -\Delta_L(x) & \mathcal{B}_L(x) & -iv_F \partial_x+\mu_L(x)  \\
\end{array}
\right),
\eeq
\beq
\mathcal{U}_{R}(x)=\left(
\begin{array}{cccc}
 -\tilde{\mu}_R(x) & \mathcal{B}_R(x) & -\tilde{\Delta}_R(x) & 0  \\
 \mathcal{B}_R(x) & -\tilde{\mu}_R(x) & 0 & -\tilde{\Delta}_R(x)  \\
 -\tilde{\Delta}_R(x) & 0 & \tilde{\mu}_R(x) & \mathcal{B}_R(x)  \\
 0 & -\tilde{\Delta}_R(x) & \mathcal{B}_R(x) & \tilde{\mu}_R(x)  \\
\end{array}
\right),
\eeq
with $\mu_L(x)=\mu\Theta(x+\frac{L}{2}),~\Delta_L(x)=\Delta\Theta(x+\frac{L}{2}),~\mathcal{B}_L(x)=\mathcal{B}\Theta(-x-\frac{L}{2}),~\tilde{\mu}_R(x)=-\mu\Theta(x-\frac{L}{2}),~\tilde{\Delta}_R(x)=-\Delta\Theta(x-\frac{L}{2}),~\mathcal{B}_R(x)=\mathcal{B}\Theta(x-\frac{L}{2})$. Similar forms can be constructed for $\mathcal{H}_R(x)$ and $\mathcal{U}_L(x)$ by noticing that $\mathcal{H}_L(x)=K\mathcal{H}_R(-x)K,~\mathcal{U}_R(x)=\mathcal{U}_L(-x),~\mu_L(x)=\mu_R(-x),~\tilde{\mu}_L(x)=\tilde{\mu}_R(-x),~\Delta_L(x)=\Delta_R(-x),~\tilde{\Delta}_L(x)=\tilde{\Delta}_R(-x),~\mathcal{B}_L(x)=\mathcal{B}_R(-x)$. Typically, $\mathcal{H}_{L,R}(x)$ and $\mathcal{U}_{R,L}(x)$ are spatially well-separated from each other, which allows to firstly approximately focus on the Majorana zero mode from only the part of $\mathcal{H}_{L,R}(x)$, then adding the treatment of $\mathcal{U}_{R,L}(x)$ as the perturbation.

\textbf{Left Majorana zero mode.} We are now in a position to find the analytical wavefunction for the normalizable Majorana bound state in the left domain wall $x=-\frac{L}{2}$ by solving the following reduced $1$D Dirac equation with the proper boundary conditions,
\begin{equation}
\left(
\begin{array}{cccc}
 -iv_F \partial_x-\mu_L(x) & \mathcal{B}_L(x) & -\Delta_L(x) & 0  \\
 \mathcal{B}_L(x) & iv_F \partial_x-\mu_L(x) & 0 & -\Delta_L(x)  \\
 -\Delta_L(x) & 0 & iv_F \partial_x+\mu_L(x) & \mathcal{B}_L(x)  \\
 0 & -\Delta_L(x) & \mathcal{B}_L(x) & -iv_F \partial_x+\mu_L(x)  \\
\end{array}
\right)\cdot\left(
\begin{array}{c}
 u(x)  \\
 w(x)  \\
 w^*(x)  \\
 -u^*(x)  \\
\end{array}
\right)=0.
\label{eq:dirac4}
\end{equation}
The self-conjugate Majorana form of the spinor here is fixed by particle-hole symmetry, defined in real space by the operator $\Xi\equiv\tau_y\otimes\sigma_yK$, where $\tau_y\otimes\sigma_y$ is the charge conjugation matrix and $K$ denotes complex conjugation. $\Xi$ anticommutes with $\mathcal{H}(x)$, and hence constrains the form of the zero-energy solutions.

In the region $x\geq-\frac{L}{2}$, solving Eq.~(\ref{eq:dirac4}) yields the $2$-component zero-mode eigenvector, $\chi_0(x)=\left(
\begin{array}{c}
 u(x) \\
 w^*(x) \\
\end{array}
\right)=\chi_{-\frac{L}{2}}\left(
\begin{array}{c}
 1 \\
 i \\
\end{array}
\right)\exp{\!\left(i\frac{\mu}{v_F}(x+\frac{L}{2})-\frac{\Delta}{v_F}(x+\frac{L}{2})\right)}$, where $\Delta>0$, and $\chi_{-\frac{L}{2}}$ is a constant. Similarly, we can solve for the wavefunction of the left Majorana zero mode in the region $x\leq-\frac{L}{2}$, where the $2$-component zero-mode eigenvector assumes $\zeta_0(x)=\left(
\begin{array}{c}
 u(x) \\
 w(x) \\
\end{array}
\right)=\zeta_{-\frac{L}{2}}\left(
\begin{array}{c}
 -i \\
 1 \\
\end{array}
\right)\exp{\!\left(\frac{\mathcal{B}}{v_F}(x+\frac{L}{2})\right)}$, with $\mathcal{B}>0$, and $\zeta_{-\frac{L}{2}}$ is a constant. The complex constants $\chi_{-\frac{L}{2}}$ and $\zeta_{-\frac{L}{2}}$ can be constrained by equating $\chi_0$ and $\zeta_0$ at $x=-\frac{L}{2}$ due to the continuity of the wavefunction: $\zeta_{-\frac{L}{2}}=-\chi_{\Im},~\chi_{-\frac{L}{2}}=i\chi_{\Im}$, where $\chi_{\Im}$ is a real constant to be fixed by the normalization condition.

The $4$-component wavefunction of the left Majorana zero mode $\Psi_L$ thus becomes $($up to normalization$)$,
\begin{equation}
\Psi_L(x)=\chi_{\Im}\left\{\left(
\begin{array}{c}
 i\exp{\!\left(i\frac{\mu}{v_F}(x+\frac{L}{2})\right)}  \\
 -\exp{\!\left(-i\frac{\mu}{v_F}(x+\frac{L}{2})\right)}  \\
 -\exp{\!\left(i\frac{\mu}{v_F}(x+\frac{L}{2})\right)}  \\
 i\exp{\!\left(-i\frac{\mu}{v_F}(x+\frac{L}{2})\right)}  \\
\end{array}
\right)e^{-\frac{\Delta}{v_F}(x+\frac{L}{2})}\Theta(x+\frac{L}{2})+\left(
\begin{array}{c}
 i  \\
 -1  \\
 -1  \\
 i  \\
\end{array}
\right)e^{\frac{\mathcal{B}}{v_F}(x+\frac{L}{2})}\Theta(-x-\frac{L}{2})\right\}.
\label{left}
\end{equation}

\textbf{Right Majorana zero mode.} Since $\mathcal{H}_L(x)=K\mathcal{H}_R(-x)K$, we can directly derive the wavefunction for the right Majorana zero mode $\Psi_R(x)=K\Psi_L(-x)=\Psi^*_L(-x)$.

\textbf{Zero-bias splitting.} If we assume $|\Phi^+_g(x)\rangle$ is the eigenstate of the full BdG Hamiltonian $\mathcal{H}(x)$ with eigenenergy $\mathcal{E}_+$, namely $\mathcal{H}(x)|\Phi^+_g(x)\rangle=\mathcal{E}_+|\Phi^+_g(x)\rangle$, then according to the particle-hole symmetry, we will have $\mathcal{H}(x)\left(\Xi|\Phi^+_g(x)\rangle\right)=-\mathcal{E}_+\left(\Xi|\Phi^+_g(x)\rangle\right)$, therefore we denote $|\Phi^-_g(x)\rangle=\Xi|\Phi^+_g(x)\rangle$, which satisfies $\mathcal{H}(x)|\Phi^-_g(x)\rangle=\mathcal{E}_-|\Phi^-_g(x)\rangle=-\mathcal{E}_+|\Phi^-_g(x)\rangle$. One can check that $\Xi|\Psi_R(x)\rangle=|\Psi_R(x)\rangle,~\Xi|\Psi_L(x)\rangle=|\Psi_L(x)\rangle$, thus we can safely assume $|\Phi^{\pm}_g(x)\rangle=\frac{1}{\sqrt{2}}\left(|\Psi_R(x)\rangle\pm i|\Psi_L(x)\rangle\right),~|\Phi^{\pm}_g(-x)\rangle=\frac{1}{\sqrt{2}}\left(|\Psi_R(-x)\rangle\pm i|\Psi_L(-x)\rangle\right)=\pm iK|\Phi^{\pm}_g(x)\rangle$.

The eigenenergies of these two approximate eigenstates are given by $\mathcal{E}_{\pm}\!\approx\!\mp~\textrm{Im}\!\left[\int^{\infty}_{0}\!\Psi^{\dagger}_R(x) \mathcal{U}_R(x) \Psi_L(x)dx\right]$, which can be further simplified: $\int^{+\infty}_{0} \Psi^{\dagger}_R(x) \mathcal{U}_R(x) \Psi_L(x)dx=iv_F \Psi^{\dagger}_R(x=0)(\tau_z\otimes\sigma_z)\Psi_L(x=0)$, where we have integrated the kinetic term by parts. Therefore, we derive: $\mathcal{E}_{\pm}\approx\mp~v_F \textrm{Re}\!\left[\Psi^{\dagger}_R(x=0)(\tau_z\otimes\sigma_z)\Psi_L(x=0)\right]$.

Thus far, our analysis is completely general, having used only the symmetries of the problem at hand. To obtain the final expression, we plug in the explicit forms for $\Psi_{L/R}$, which gives:
\beq
E_{\pm}=\frac{1}{2}\mathcal{E}_{\pm}\approx\pm~2v_F \chi^2_{\Im}~e^{-\frac{\Delta L}{v_F}}\cos{\!\left(\frac{\mu L}{v_F}\right)}=\pm~\frac{\Delta\mathcal{B}}{\Delta+\mathcal{B}}~e^{-\frac{\Delta L}{v_F}}\cos{\!\left(\frac{\mu L}{v_F}\right)}\xrightarrow{\mathcal{B}\gg\Delta}\pm~\Delta~e^{-\frac{\Delta L}{v_F}}\cos{\!\left(\frac{\mu L}{v_F}\right)}.
\label{eq:eplusminus}
\eeq
A parallel evaluation can be worked out for the SC-FM-SC setup with the inclusion of a finite magnetic field term.

It is instructive to compare this result with that obtained in the main text. Importantly, we find that the oscillatory cosine term has exactly the same form in both calculations, and vanishes if the chemical potential is tuned to zero. However, they differ in the exponential decay rate. Let us summarize the key result of this section: the zero-energy splitting in the quantum spin-Hall edge has an oscillatory dependence on $\mu$, which exactly matches that found in the instanton calculation.

\section{Instantons in finite-interval sine-Gordon model with Neumann boundary conditions} \label{SupplementaryInst}

In the instanton calculation presented in the main text, a crucial element is the evaluation of the Fredholm determinant. For periodic boundary conditions, this has been discussed for the sine-Gordon theory in Ref.~\cite{Bajnok}; here we will review the relevant aspects of the derivation, and explain how the situation differs for open boundaries.

To evaluate the functional determinant, we must first explicitly factor out the zero-mode resulting from time-translation of the center of the instanton. This leads to a factor of $\sqrt{S_0/(2\pi\hbar)}$ $($see, e.g. Ref.~\cite{Vainshtein}$)$. In addition, one typically evaluates a ratio of the functional determinant of interest to the functional determinant of a harmonic oscillator, including the latter into the normalization factor.

Using the dilute instanton gas approximation, in this approach formally the energy splitting is given by \cite{Bajnok}
\begin{align}
E(q_{\phi})&=-2\hbar\sqrt{v}\cos{\!\left(\frac{\pi}{n}q_{\phi}-\frac{\mu L}{\hbar nv}\right)}\cdot\left(\sqrt{\frac{S_0}{2\pi\hbar}}\cdot e^{-S_0/\hbar}\right)\cdot\left(\frac{\sqrt{\det{\left[-\partial^2_{x'}-\partial^2_{\tau'}+\frac{4\Delta}{\hbar\varsigma}\right]}}}{\sqrt{\det'\left[-\partial^2_{x'}-\partial^2_{\tau'}-\frac{4\Delta}{\hbar\varsigma}\sin{\!(2n\phi_{sol}(\tau))}\right]}}\right), \label{Esplit0}
\end{align}
where we have rescaled the variables and the fields, and the prime on $\det'$ means that the zero eigenvalue has been excluded.

In a one-dimensional system, the ratio of determinants in Eq.~(\ref{Esplit0}) is divergent due to spatial fluctuations, and must be regulated. As shown in Ref.~\cite{Munster}, the divergent contribution to the functional determinant ratio is canceled by using renormalized values of the effective mass and coupling constant in $S_0$, and the finite part can be expressed:
\beq
\log{\!\left(\frac{\det'\hat{M}}{\det\hat{M}_0}\right)}\longrightarrow-\text{lim}_{z\rightarrow 0} \frac{d}{dz}\zeta(z,\hat{M})=-\frac{d}{dz}\left(\frac{1}{\Gamma(z)}\int^{\infty}_0 dt~t^{z-1}\{K_t(\hat{M})-K_t(\hat{M}_0)-1\}\right)\!\Big\vert_{z\rightarrow0}.
\label{divint}
\eeq
Here the heat kernel $K_{t}(\hat{A})=\textrm{Tr}{[e^{-t\hat{A}}]}$ and the Hermitian operators $\hat{M},~\hat{M}_0$ and $\hat{Q},~\hat{Q}_0$ are defined as follows,
\begin{align}
\hat{M}&=-\partial^2_{x'}-\partial^2_{\tau'}+m^2_0\left(1-\frac{2}{\cosh^2\left(m_0(\tau'-\tau'_0)\right)}\right)=-\partial^2_{x'}+\hat{Q},~~~\hat{M}_0=-\partial^2_{x'}-\partial^2_{\tau'}+m^2_0=-\partial^2_{x'}+\hat{Q}_0, \label{Mdef}
\end{align}
where the mass term $m_0=\sqrt{4\Delta/(\hbar\varsigma)}$.

It is useful to separate the zeta function into two main pieces: $\zeta(z,\hat{M})=\zeta_{\textrm{log}}(z,\hat{M})+\zeta_{\textrm{mix}}(z,\hat{M})$, where
\begin{align}
\zeta_{\textrm{log}}(z,\hat{M})&=\frac{1}{\Gamma(z)}\int^{\infty}_0 dt~t^{z-1}\left\{K_t(-\partial^2_{x'})-1\right\},~~\zeta_{\textrm{mix}}(z,\hat{M})=\frac{1}{\Gamma(z)}\int^{\infty}_0 dt~t^{z-1}\left\{K_t(-\partial^2_{x'})\cdot\left[K_t(\hat{Q})-K_t(\hat{Q}_0)-1\right]\right\}, \nonumber
\end{align}
with $\hat{Q}, \hat{Q}_0$ given in Eq.~(\ref{Mdef}).

Of primary interest is the contribution of $\zeta_{\textrm{log}}$, which gives rise to the extra $1/\sqrt{L}~\left(L=\sqrt{v}L'\right)$ factor in the prefactor of the energy splitting, leading to an $L$-independent prefactor in agreement with known results for Majorana zero modes \cite{DasSarma,Zyuzin}. This term can be computed exactly:
\begin{align}
\zeta_{\textrm{log}}(z,\hat{M})&=\frac{1}{\Gamma(z)}\int^{\infty}_0 dt~t^{z-1}\{K_t(-\partial^2_{x'})-1\}=\sum^{\infty}_{n=1}\frac{1}{\Gamma(z)}\int^{\infty}_0 dt~t^{z-1}e^{-k^2_nt}=\sum^{\infty}_{n=1}\left(\frac{1}{k^2_n}\right)^z. \nonumber
\end{align}
In the continuum limit with the von Neumann boundary conditions, the momentum $k_n=\frac{n\pi}{L'},~n=0,1,2,\ldots,$ so
\beq
\zeta_{\textrm{log}}(z,\hat{M})=\left(\frac{L'^2}{\pi^2}\right)^z\sum^{\infty}_{n=1}\left(\frac{1}{n^2}\right)^z=\left(\frac{L'^2}{\pi^2}\right)^z\zeta(2z)~~~\mbox{and}~~~\exp{\!\left(\frac{1}{2}\frac{d\zeta_{\textrm{log}}}{dz}\Big\vert_{z=0}\right)}=\frac{1}{\sqrt{2}}\frac{1}{\sqrt{L'}}=\frac{v^{\frac{1}{4}}}{\sqrt{2}}\frac{1}{\sqrt{L}}. \nonumber
\eeq
Crucially, here we have kept only positive $k$ values in the sum, as is appropriate for general $($i.e. fixed or free$)$ open boundary conditions. For periodic boundary conditions, where both positive and negative $k$ are allowed, $\zeta_{\textrm{log}}$ is proportional to $1/L$ \cite{Munster}.

The remaining term $\zeta_{\textrm{mix}}$ is important chiefly because, for open boundary conditions, it contributes a dimensionful constant to the prefactor. It is convenient to divide this into the following three contributions:
\begin{align}
\zeta_{\textrm{mix}}(z,\hat{M})&=\zeta_{\textrm{con}}(z,\hat{M})+\zeta_{\textrm{lin}}(z,\hat{M})+\zeta_{\textrm{exp}}(z,\hat{M}), \nonumber \\
\zeta_{\textrm{con}}(z,\hat{M})&=\frac{1}{\Gamma(z)}\int^{\infty}_0 dt~t^{z-1}\left\{\frac{1}{2}\cdot\left[K_t(\hat{Q})-K_t(\hat{Q}_0)-1\right]\right\}, \nonumber \\
\zeta_{\textrm{lin}}(z,\hat{M})&=\frac{1}{\Gamma(z)}\int^{\infty}_0 dt~t^{z-1}\left\{\frac{L'}{4\sqrt{4\pi t}}\cdot\left[K_t(\hat{Q})-K_t(\hat{Q}_0)-1\right]\right\}, \nonumber \\
\zeta_{\textrm{exp}}(z,\hat{M})&=\frac{1}{\Gamma(z)}\int^{\infty}_0 dt~t^{z-1}\left\{\left(K_t(-\partial^2_{x'})-\frac{1}{2}-\frac{L'}{4\sqrt{4\pi t}}\right)\cdot\left[K_t(\hat{Q})-K_t(\hat{Q}_0)-1\right]\right\}. \nonumber
\end{align}
$\zeta_{\textrm{exp}}(z,\hat{M})$ gives an additive contribution that is exponentially suppressed in $L$ relative to the other two terms, and can be neglected. We will show that the $\zeta_{\textrm{con}}$-contribution -- which is not present with periodic boundary conditions -- ensures the correct dimensionality for the splitting.

To compute these, we require the spectra and densities of states of the operators $\hat{Q}$ and $\hat{Q}_0$. Apart from the zero-mode of $\hat{Q}$, which we have already explicitly factored out, these are continuous functions, and are given in Ref.~\cite{Bajnok}. This gives:
\begin{align}
\zeta_{\textrm{con}}(z,\hat{M})&=\frac{1}{\Gamma(z)}\int^{\infty}_0 dt~t^{z-1}\left\{\frac{1}{2}\cdot\left[-\frac{1}{\pi}\int^{\infty}_{0}dp\left(\frac{2m_0}{p^2+m^2_0}\right)e^{-t(p^2+m^2_0)}\right]\right\}=-\frac{1}{2\sqrt{\pi}}\cdot m^{-2z}_0\cdot\frac{\Gamma(z+1/2)}{\Gamma(z+1)}. \nonumber
\end{align}
Therefore, the extra contribution to the prefactor assumes
\beq
\frac{1}{2}\frac{d\zeta_{\textrm{con}}(z,\hat{M})}{dz}\Big\vert_{z=0}=\frac{1}{2}\log{\left(2m_0\right)}~~~\mbox{and}~~~\exp{\!\left(\frac{1}{2}\frac{d\zeta_{\textrm{con}}(z,\hat{M})}{dz}\Big\vert_{z=0}\right)}=\sqrt{2m_0}=\sqrt{2}\left(\frac{4\Delta}{\hbar\varsigma}\right)^{\frac{1}{4}}=\frac{\sqrt{2\omega}}{v^{\frac{1}{4}}},
\label{zetacon}
\eeq
where we use the definition of $\omega=2\sqrt{\Delta v/(\hbar\varsigma)}=m_0\sqrt{v}$ in the main text. This ensures that the overall prefactor has the dimensions of energy. For periodic boundary conditions, where the contribution of $\zeta_{\textrm{log}}\sim 1/L$, there is no $\zeta_{\textrm{con}}$-contribution.

$\zeta_{\textrm{lin}}$ can be calculated in a similar manner, and makes an additive contribution to $S_0$ that scales linearly with $L$. This must be combined with other such finite terms that arise in fixing the renormalized value of $m$ relative to the bare value appearing in $S_0$. $($Recall that it is necessary to use this renormalized value, as it cancels the divergences in the determinant ratio in Eq.~(\ref{Esplit0})~\cite{Munster}$)$. Ref.~\cite{Bajnok} verified numerically that the net effect of these finite contributions on the energy splitting is negligible for the sine-Gordon model with periodic boundary conditions; since these contributions are independent of the boundary conditions, we expect that this holds in the present context as well, and leave a more precise evaluation of the leading quantum corrections to $S_0$ for the future investigations.

The final finite result for the energy splitting thus assumes the following form,
\beq
E(q_{\phi})=-2\hbar\sqrt{\frac{v\omega}{L}}\sqrt{\frac{S_0}{2\pi\hbar}}\cos{\!\left(\frac{\pi}{n}q_{\phi}-\frac{\mu L}{\hbar nv}\right)}\cdot\exp{\!\left(-\frac{2\omega}{n\pi v}L\right)}=-\frac{2\hbar\omega}{\pi}\sqrt{\frac{1}{n}}e^{-S_0/\hbar}\cos{\!\left(\frac{\pi}{n}q_{\phi}-\frac{\mu L}{\hbar nv}\right)}.
\eeq
After incorporating both the spatial and temporal fluctuations, the overall dependence of the prefactor on the system's size becomes independent of $L$, which results from a cancelation between the factor of $\sqrt{S_0}$ due to the time-translational-invariant zero-mode, and the effect of finite-momentum fluctuations in the Fredholm determinant with the Neumann boundary conditions.

\end{widetext}


\begin{thebibliography}{99}
\bibitem{FootnoteAdd3} It is worth noticing that the chiral phases are not allowed in the Ising chain.
\end{thebibliography}
\end{document}